Anisotropic polaritons in 2D vdW materials

A. Babar Shabbir[a], B. Weiliang Ma[b] and C. Qiaoliang Bao[a*].

[a] Department of Materials Science and Engineering, Monash University, Clayton, Victoria 3800, Australia

[b] Wuhan National Laboratory for Optoelectronics and School of Optical and Electronic Information, Huazhong University of Science and Technology, Wuhan 430074, China

*corresponding email address: qiaoliang.bao@gmail.com

ABSTRACT

Perhaps the most significant progress to the field of infrared optics and nanophotonics has been made through the real space realisation of polaritons in two-dimensional materials that provide maximum light confinement functionalities. The recent breakthrough discovery of in-plane hyperbolicity in the natural van der Waals material has revealed a most exciting optical property which enable an in-plane anisotropic dispersion. Yet, the most intriguing feature of in-plane anisotropic dispersion is the manipulation of polaritons at the nano scale. This development has opened a new window of opportunity in order to develop unique nanophotonic devices with unprecedented controls. This chapter will cover these developments with focus on fundamental understandings and progress of real space visualisation of in-plane anisotropic polaritons in the near-field range. The last section will conclude with the future prospects of this rapidly emerging area.

**X.1 Introduction**

Optoelectronics and photonics support our modern day lives by enabling communication, transport, computing and smart phones, among many other things. Achieving these modernized "basic" functions, has been a key driver for technological innovation and the formation of a global economic engine. Presently, the photonics market alone is worth more than USD$500 billion worldwide and is predicted to grow to over USD$600 billion by 2023, while the combined market caps of the two largest US smart-phone manufactures increased from US$ 575 billion in 2010 to US$2.3 trillion in 2019 [The most valued technology gadgets of the last decade, Deloitte 2020]. To highlight the major job and global impacts that these sectors have, the global biophotonics market is expected to be worth only USD$50 billion by the end of 2020, according to the London-based "Report Buyer" market research services. The optoelectronics and photonics industries are now placing more emphasis on developing new light weight and energy efficient/dissipationless technologies based on advanced materials in response to increasingly high-tech demands and end user challenges. Exploiting polaritons in such applications may prove to be the answer.

Polaritons (a hybrid of light-matter oscillations) are bosonic quasiparticles formed by the strong coupling of photons with an electric or magnetic dipole-carrying excitation in van der Waals materials (vdW). So far, different types of polaritons in vdW crystals, including those formed by free electrons (plasmon polaritons),[1-5] strongly-bound electron-hole pairs or excitons (exciton-polaritons)[6-8] and bound lattice vibrations or phonons (phonon polaritons)[9-11] have been experimentally discovered and already shown a numerous potentials in light-based technologies beyond sub-diffraction limit.[12, 13] There are also predictions of Cooper-pair polaritons (formed due to a coupling of photons with cooper-pairs in vdW superconducting materials), and magnon polaritons (due to a coupling of photons with magnetic resonances in vdW magnetic materials). The highly confined phonon polaritons with hyperbolicity attract growing interest for their capability of manipulating light at the nanoscale because of their strong field confinement, low losses and long lifetimes.[10]

The propagating nature of light inside a medium can be explained by the dielectric permittivity tensor i.e. $\varepsilon = \text{diag}\,[\varepsilon_x, \varepsilon_y, \varepsilon_z]$ at such frequency, where $\varepsilon_x$, $\varepsilon_y$ represents the in-plane components and $\varepsilon_z$ indicate the out-of-plane component.[14] The unusual properties of propagating light inside a medium are linked to the isofrequency surface in the momentum space, which can be described by the sign of $\varepsilon_i (i = x, y, z)$. In case of crystallographically anisotropic medium, different dielectric responses can be observed along different directions. For example, when one of the permittivities, *i.e.*, $\varepsilon_i (i = x, y, z)$, is opposite in sign to the other two components, such medium can therefore, be regarded as a hyperbolic medium.[15, 16] The hyperbolic materials are essential to a range of sub-diffraction optical features , such as negative refraction,[17, 18] hyperlens,[19-21] nanolithography,[22] and enhancement of quantum radiation.[23] Also, the hyperbolicity can also be found in planar surface when the condition i.e. $\varepsilon_x * \varepsilon_y < 0,$ is satisified.[24] In simple words, hyperbolic materials are extremely reflective to light along a specific axis and normal reflective to light along an orthogonal axis. Generally, both these axes are in different planes for example one of these axes lie in the plane and whereas the other axis could be out of the plane.

Assuming the different permittivities along in-plane orthogonal directions, the polaritons with an in-plane anisotropic dispersion can propagate along the layers. The polaritons can also display an in-plane elliptical dispersion due to different permittivities however, with the same signs while; for opposite signs, the polaritons can propagate with an in-plane hyperbolic dispersion. Take hexagonal boron nitride (hBN) as an example; which is a natural hyperbolic material.[20] The hBN is a polar vdW material with uniaxial permittivities holding two mid-infrared Reststrahlen bands, generally known as type 1 or lower (~760 < $\omega$ < 825 cm$^{-1}$) and type II or upper (~1395 to 1630cm$^{-1}$) bands for enriched hBN. In type II; the in-plane permittivity is negative and shows an isotropic nature (i.e. $\varepsilon_x = \varepsilon_y = \varepsilon_\perp < 0$); whereas; the out of plane permittivity is positive (i.e. $\varepsilon_z = \varepsilon_\parallel > 0$).[21]

As a result, the phonon polaritons in natural hBN exhibit an out-of-plane hyperbolic dispersion and isotropic in-plane dispersion. Even though the phonon polaritons with an in-plane hyperbolic dispersion can be observed in artificially created hBN nanostructures but their real space visualisation in natural materials were highly desirable because of its useful applications to the polaritonic technologies.[25] The recent discovery of anisotropic polariton propagation along the surface of a natural α-MoO$_3$ vdW flakes has revealed the existence of such elliptic and hyperbolic in-plane dispersions.[26] This has generated huge interests in the fields of infrared optics and nanophotonics because natural vdW materials with in-plane hyperbolic dispersions could help us to manipulate light with unprecedented controls and particularly, with extreme subdiffractional confinement, ultra-low losses, and polarization sensitivities.

This book chapter will give an overview of the recent progresses in the field of hyperbolic materials with an emphasis on in-plane hyperbolic natural materials. Following sections are carefully designed so as to enable the readers to easily understand the basics concepts, related aspects and the latest progress in this field. First section will describe the real-space nanoimaging mechanism utilising scattering-type scanning near-field optical microscopy (s-SNOM) and outline the overall research progress in polaritonic field. The basic concepts of anisotropic polaritons with hyperbolic and/or elliptic dispersions will be explained. Other sections will cover hyperbolic metasurfaces and the in-plane anisotropic natural materials. Last section will discuss several possible applications of this exciting but emerging field.

**X.2 Real space imaging of polaritons in vdW Crystals**

Polaritons in vdW crystals can help us to manipulate light at nanoscale, which is much smaller than the sub-diffraction limit.[27] Although, this was initially achieved by metal plasmonics[28-30] but the related losses are huge due to electron/plasma scattering and interband transitions.[31, 32] The emergence of 2D vdW materials provide a required platform to discover relatively low-loss polaritons for nanophotonic applications. Meanwhile, the invention of the s-SNOM can help us to directly visualise these highly confined polaritons in the near field range, which can help us to probe the fundamental physics of low-dimensional materials. The s-SNOM provides a wavelength-independent nanoscale spatial resolution (~10 nm, determined by the diameter of the tip apex), and thus enabling us to characterise different polaritons.[33] This equipment also demonstrate its powerful capabilities to explore several quantum phenomenas associated to polaritons.[34] In its working conditions, light with required wavelengths are focused onto a high-frequency oscillating metallic tip of an AFM (i.e. atomic force microscope). The tip acts as an optical antenna that effectively produces optical near fields with wave vectors much larger than that of the illuminating light. This can match the excitation conditions of the polaritons in the materials of interest. Using s-SNOM recorded real space images, the polariton wavelength and propagating properties can be estimated from the local optical field and the obtained periodicity in the interference fringes. The dispersion relation and propagation losses from the real-space imaging can therefore be calculated. For further reading on s-SNOM technique topic in details, see a review paper on Modern Scattering-Type Scanning Near-Field Optical Microscopy for Advanced Material Research by Chen *et al.*[33] The hyperspectral imaging by the nano-FTIR (nanoscale Fourier transform infrared spectroscopy) can also provide the spectral and spatial resolution in nanometer scale to further investigate the polaritonic behaviours as the metallic tip is raster-scanned over the sample (line-scans). The frequency dependence of the nano-FTIR spectral images can provide additional insights into the polaritonic properties and its combination with s-SNOM can depict a complete picture of understanding polaritons.[35, 36]

In 2011, Chen *et al.*[12] and Fei *et al.*[1] (independently) used s-SNOM for real space imaging of the polaritonic waves in graphene for the first time and reported the strong light confinement and long lifetime of nearly 2ps. The plasmon polariton in in such an atomically thin layer are in-plane isotropic and can be tuneable by approaches such as gating, doping, dielectric environment and photoexcition.[2, 3, 37, 38] This innovation opens a new avenue of research, which also inspired the others to investigate various form of polaritons in different 2D materials[39-44], and utilising polaritons into low-loss gate-tuneable nanophotonic devices.

As a result, the hyperbolic phonon polaritons in hBN were discovered by Dai and co-authors in 2014.[45] Depending on material's thickness, these phonon polaritons can be visualised down to few layers as well as monolayer hBN.[46] The hyperbolic phonon polariton in hBN can exist as volume confined modes as these hyperbolic modes are localised inside extremely small volumes. This could be connected to the presence of large birefringence between in plane and out of plane. Due to the volume-confined modes within hBN, the development of unique functionalities such as the nanoscale equivalent of an optical fibre could potentially be possible. The hyperbolic phonon polaritons are in-plane isotropic and can be tuneable by utilising several methods including thickness, dielectric environment and isotope enrichment.[9, 47-49] These polaritons can propagate with nearly 8ps lifetime.

Special attention was given to the fabrication of new vdW heterostructures as this can offer an extra degree of freedom to tune these polaritons and that can provide new perspectives in this rapidly emerging field. Take graphene/hBN heterostructure as an example; which can support a unique hybrid of plasmon-phonon polariton. These hybrid modes have long lifetimes which can be easily controlled by electric gating.[37, 50-52] Later on, Fei[7] and Hu[6] *et al.* visualised the real-space propagation of exciton polariton waveguide modes in vdW $WSe_2$ and $MoSe_2$ at room temperatures. The nanoimaging of exciton-polaritons demonstrates the high sensitivity of the polariton propagation length with respect to the excitation energy. The obtained large propagation length ~12 µm in exciton polariton could be useful to long-range energy or information transfer in future nanophotonic circuit technologies. Also, the exciton polariton in $WSe_2$ and $MoSe_2$ are in-plane isotropic in nature and they depend on material's thickness and dielectric environment.[6, 7, 53]

Furthermore, the metasurface (i.e. a rationally designed artificial interface, with subwavelength thickness and that can manipulate light by spatially arranged meta-atoms generally known as building blocks of the metasurface) based on 2D materials particularly in, with extreme anisotropic optical properties, which can support polaritons were investigated to study peculiar quantum properties associated to these vdW structures. The realization of in-plane anisotropic polaritons in both natural α-MoO₃ and nanostructured hBN metasurface indicate a great opportunity to manipulate infrared light with highly directional controls. The ultralow loss in-plane anisotropic polaritons in natural α-MoO₃ can propagate with lifetime of over 8ps and can be tuneable by material's thickness and metal-atom intercalation approaches.[26, 54] Researchers also used Moire polaritons to uncover the secrets of polaritonic behaviours in 2D materials. This has resulted in the observation of surface plasmons with unprecedented control and tunability in moire graphene superlattice,[55] moire exciton polariton in twisted stacks of monolayer transition metal dicalcogenides,[56] phonon-based soliton superlattices in twisted stacks of hBN,[57] and topological polaritons and photonic magic angles in twisted α-MoO3 bilayers.[58]

Here, it is important to mention BP, a promising novel material in polariton research area and having advantages of ultrafast switching. Huber and co-authors designed a SiO₂/black phosphorus/SiO₂ heterostructure in which the surface phonon modes of the SiO₂ layers hybridize with surface plasmon modes in black phosphorus (BP).[3] These in-plane anisotropic plasmon polaritons in BP can be activated by photo-induced interband excitation.

**X.3 Anisotropic polaritons in 2D vdW materials**

The nature of propagating light inside a medium is related to a iso-frequency contour in the *k*-space. The momentum *k* corresponding to each polaritonic wavelength can be described by permittivities at constant frequencies (*ω)*. Therefore, permittivity is a frequency-dependent tensor and defined as $\varepsilon = \text{diag}\,[\varepsilon_x, \varepsilon_y, \varepsilon_z]$, here $\varepsilon_x$, $\varepsilon_y$ indicates the in-plane components and $\varepsilon_z$ represent the out-of-plane component. The medium is illustrated as hyperbolic material when the permittivity tensor along different orthogonal axes are of opposite signs.[59] In plain words, hyperbolic materials are extremely reflective to light along a specific axis and normal reflective to light along an orthogonal axis. Generally, both these axes are in different planes for example one of these axes lie in the plane and whereas the other axis could be out of the plane.

The materials with extreme optical anisotropies are useful in plenty of nanophotonic applications such as negative refraction, hyperlens, enhanced quantum radiation, and etc.[60-64] Initially, the hyperbolic media was generally explored in metamaterials (i.e. artificially designed materials; with unique properties usually gained from their structures and engineered unit cells or meta-atom, rather than from the properties of their constitutive materials ), aiming at the discoveries of unique properties, which are hard to find in in natural materials. The realization of the hyperbolicity in natural hBN has open up new avenues for sub-diffraction imaging and focusing, exploring ultralow loss polaritonic propagations and their associated quantum phenomena.[65] The bulk hyperbolicity has also been found in other class of natural materials such as graphite,[66] magnesium diboride,[67] quartz,[68] tetradymites,[69] calcite,[70] and α-$Al_2O_3$ single crystal.[71] Usually the polaritons supported by these mediums are in-plane isotropic as both axis (which are highly reflective to light and normal reflective to light, respectively) are in different planes i.e. in plane and out of plane. However, realizing both these axis in a same plane is challenging (i.e. in plane anisotropic polaritons), which are highly desirable for nanopolaritonic applications such as imaging and energy transfer devices.[72, 73]

As mentioned earlier, the wave propagation in in-plane anisotropic medium is related to an iso-frequency contour in *k*-space, which can also be linked to the Fourier transformation of the propagating wave in real space. To differentiate the different forms of in-plane anisotropic media, the in-plane iso-frequency contour in the wave vector space is shown Figure X.1, which also illustrate the in-plane iso-frequency contour and its dependence on the relative sign of the real component of permittivities Re(ε) or the imaginary component of conductivities Im(σ). Together, the surface plasmon polariton in graphene,[74] hyperbolic phonon polaritons in hBN,[75] surface phonon polaritons in SiC[76] demonstrate the in-plane isotropic polaritons and illustrated as a circular spreading in real space (see Figure X.1a). The symmetrical propagation of polaritons can therefore, not be utilised for applications such as in-plane energy transfer or polarization.

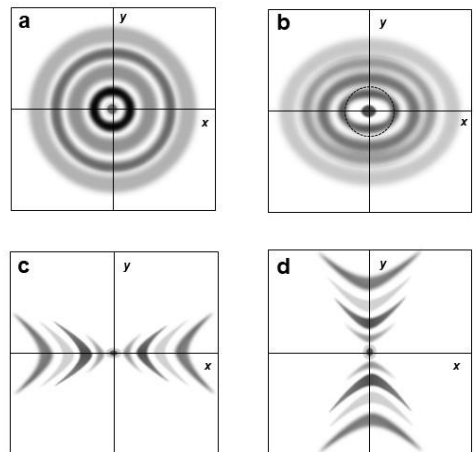

Figure X.1 Field distribution excited by a z-directed dipole for an; (a) in-plane isotropic media. (b) in-plane elliptic media. (i.e. $\epsilon_x * \epsilon_y > 0, \epsilon_x \neq \epsilon_y$). (c) in-plane hyperbolic media. (i.e. $\epsilon_x < 0, \epsilon_y > 0$). (d) in-plane hyperbolic media. (i.e. $\epsilon_x > 0, \epsilon_y < 0$)

On the other side, the elliptical and hyperbolic curves of in-plane iso-frequency contours (see Figure X.1c-d) suggest an anisotropic propagation of polaritons with diverse *k* along different directions. It is important to note here that the propagation along both *x*- and *y*- directions are allowed in the elliptic case, whereas; the propagation along one of the axes is not permitted in case of hyperbolic geometry. Also, the huge optical anisotropy will contribute in-plane hyperbolicity (i.e. Figure X.1c

& d) and the large-k polaritons are linked to open curve in momentum space, which could be used as a platform to enhance quantum radiation. The nanophotonic devices based on them can easily be integrated to the other flat optical and optoelectronic elements due to their planar nature i.e. in-plane.

## X.4 Dispersion Relation for in-plane Anisotropic Polaritons

The focus of this book chapter is to closely look the all aspects of in-plane anisotropic polaritons and therefore, this section will cover the dispersion relation for α-MoO$_3$. The dispersion relation for anisotropic polaritons in vdW crystals was calculated by Bao and colleagues[26] by assuming α-MoO$_3$ flake as a 2D conductivity sheet of zero thickness (i.e. $z = 0$). The flake is supposed to be placed between two dielectric half-spaces with the permittivities $\varepsilon_1$ (region "1", $z > 0$) and $\varepsilon_2$ (region "2", $z < 0$).

The conductivity tensor can be written as:

$$\hat{\sigma} = \begin{pmatrix} \sigma_{xx} & 0 \\ 0 & \sigma_{yy} \end{pmatrix} \quad (1)$$

The fields above ($z > 0$) and below ($z < 0$) the sheet can be represented by utilising the following polarization basis vectors for the electric fields in the form of Dirac notations:

$$|s1,2\rangle = \frac{1}{k_t}\begin{pmatrix} -k_y \\ k_x \\ 0 \end{pmatrix} e^{ik_x x + ik_y y}, \quad |p1,2\rangle = \frac{1}{k_t}\begin{pmatrix} k_x \\ k_y \\ \frac{k_t^2}{\mp k_{z1,2}} \end{pmatrix} e^{ik_x x + ik_y y} \quad (2)$$

The $k_x$, $k_y$ denotes the in-plane wavevector components; where $k_t^2 = k_x^2 + k_y^2$. The $k_{z1,2} = \sqrt{\varepsilon_{1,2} k_0^2 - k_x^2 - k_y^2}$ represents an out-of-plane wavevector component and $k_0 = \frac{\omega}{c}$. The $|s1,2\rangle$ and $|p1,2\rangle$ are the s- and p-polarization vectors, respectively. The - (+) signs in $|p1,2\rangle$ defination represents the wave propagation along (opposite to) the z-axis, respectively, that is in the half-space 1 (half-space 2), respectively. The subvectors (corresponds to vectors in equation 2 to illustrate the in-plane components of the fields are described as:

$$|s\rangle = \frac{1}{k_t}\begin{pmatrix}-k_y\\k_x\end{pmatrix}e^{ik_xx+ik_yy}, \quad |p\rangle = \frac{1}{k_t}\begin{pmatrix}k_x\\k_y\end{pmatrix}e^{ik_xx+ik_yy} \quad (3)$$

On the other side, the fields of the polariton in the upper and lower media can be shown as:

$$\vec{E}_1 = \sum_\beta a_\beta^+ |\beta 1\rangle e^{ik_{z1}z}, \quad \vec{E}_2 = \sum_\beta a_\beta^- |\beta 2\rangle e^{-ik_{z2}z}, \quad (4)$$

The $\beta$ indicates polarizations $s, p$, and $a_\beta^+, a_\beta^-$ are the unknown amplitudes. According to the Maxwell equations, the magnetic fields above and below the conducting sheet can be written as:

$$\vec{H}_1 = \vec{q}_1 \times \vec{E}_1, \quad \vec{H}_2 = \vec{q}_2 \times \vec{E}_2, \quad (5)$$

here $\vec{q}_{1,2}$ represents a normalized wavevector, $\vec{q}_{1,2} = \frac{\vec{k}_{1,2}}{k_0}$, $\vec{k}_{1,2} = (k_x, k_y, \pm k_{z1,2})$, where in the last expression, 1 and 2 correspond to the signs + and -, respectively. We can write the boundary conditions at the conducting sheet ($z = 0$) as;

$$\vec{E}_{1t} = \vec{E}_{2t}, \quad (6)$$

$$\vec{e_z} \times (\vec{H}_1 - \vec{H}_2) = 2\hat{\alpha}\vec{E}_{1t}. \quad (7)$$

here a normalized conductivity tensor $\hat{\alpha} = \frac{2\pi}{c}\hat{\sigma}$ is implied to simplify the equation. In equations (6) and (7); the subscript "t" indicates the in-plane components. By utilising equations (4) & (5); the equations (6) & (7) can be re-written as;

$$\sum_\beta a_\beta^+ |\beta\rangle = \sum_\beta a_\beta^- |\beta\rangle, \quad (8)$$

$$\sum_\beta \left(a_\beta^+ \vec{e}_z \times \vec{q}_1 \times |\beta 1\rangle - a_\beta^- \vec{e}_z \times \vec{q}_2 \times |\beta 2\rangle\right) = \sum_\beta 2 a_\beta^+ \hat{\alpha} |\beta\rangle. \quad (9)$$

We can simplify equation (9) by employing;

$$\vec{e}_z \times \vec{q}_1 \times |\beta 1\rangle = -Y_\beta^+ |\beta\rangle, \text{ and } \vec{e}_z \times \vec{q}_2 \times |\beta 2\rangle = -Y_\beta^- |\beta\rangle \quad (10)$$

Here $Y_\beta^\pm$ being the admittances for s- and p-polarized waves with propagation along or opposite to the z-axis:

$$Y_s^\pm = \pm q_{z1,2}, \quad Y_p^\pm = \pm \frac{\varepsilon_{1,2}}{q_{z1,2}} \quad (11)$$

Considering the expressions as mentioned in Equation (10), equation (9) can be rewritten as:

$$-\sum_\beta (a_\beta^+ Y_\beta^+ |\beta\rangle - a_\beta^- Y_\beta^- |\beta\rangle) = 2 \sum_\beta a_\beta^+ \hat{\alpha} |\beta\rangle \quad (12)$$

Multiply both sides of equation (12) by $\langle \beta |$, in which the exponential should be complex conjugated (scalar products), we can have therefore;

$$-\sum_{\beta'} (a_{\beta'}^+ Y_{\beta'}^+ \langle \beta|\beta'\rangle - a_{\beta'}^- Y_{\beta'}^- \langle \beta|\beta'\rangle) = 2 \sum_{\beta'} a_{\beta'}^+ \langle \beta|\hat{\alpha}|\beta'\rangle \quad (13)$$

Since $\langle \beta|\beta'\rangle = \delta_{\beta,\beta'}$ (Kronecker symbol), therefore;

$$-\left(a_\beta^+ Y_\beta^+ - a_\beta^- Y_\beta^-\right) = 2 \sum_{\beta'} a_{\beta'}^+ \langle \beta|\hat{\alpha}|\beta'\rangle \quad (14)$$

Since $a_\beta^+ = a_\beta^-$ (i.e. the equality of the amplitudes above and below the conducting sheet) when multiplying equation (8) with $\langle \beta|$. we can therefore reduce equation (14) to (neglect the notation signs of $a_\beta^+$ and $a_\beta^-$ due to equality);

$$-a_\beta \left(Y_\beta^+ - Y_\beta^-\right) = 2 \sum_{\beta'} a_{\beta'} \langle \beta|\hat{\alpha}|\beta'\rangle = 2 \sum_{\beta'} a_{\beta'} M_{\beta\beta'}, \quad (15)$$

Eq. (15) can be expanded as:

$$\begin{cases} \left(M_{pp} + \frac{Y_p^+ - Y_p^-}{2}\right) a_p + M_{ps} a_s = 0 \\ \left(M_{ss} + \frac{Y_s^+ - Y_s^-}{2}\right) a_s + M_{sp} a_p = 0 \end{cases}, \quad (16)$$

We can use the following matrix $\widehat{M}$ for compactness:

$$M_{\beta\beta'} = \langle \beta | \hat{\alpha} | \beta' \rangle \quad (17)$$

The system (16) has nontrivial values only when its determinant equals to zero;

$$\left(M_{pp} + \frac{Y_p^+ - Y_p^-}{2}\right)\left(M_{ss} + \frac{Y_s^+ - Y_s^-}{2}\right) - M_{sp} M_{ps} = 0 \quad (18)$$

Equation (18) constitutes the dispersion relation for polaritons. To find the explicit expression for the elements of the matrix $\widehat{M}$ we can write:

$$\left.\begin{aligned}
\langle p|\hat{\alpha}|p\rangle &= \frac{1}{k_t^2}(k_x \quad k_y)\begin{pmatrix}\alpha_{xx} & 0 \\ 0 & \alpha_{yy}\end{pmatrix}\begin{pmatrix}k_x \\ k_y\end{pmatrix} = \frac{k_x^2 \alpha_{xx} + k_y^2 \alpha_{yy}}{k_t^2} \\
\langle p|\hat{\alpha}|s\rangle &= \frac{1}{k_t^2}(k_x \quad k_y)\begin{pmatrix}\alpha_{xx} & 0 \\ 0 & \alpha_{yy}\end{pmatrix}\begin{pmatrix}-k_y \\ k_x\end{pmatrix} = \frac{-k_x k_y \alpha_{xx} + k_y k_x \alpha_{yy}}{k_t^2} \\
\langle s|\hat{\alpha}|p\rangle &= \frac{1}{k_t^2}(-k_y \quad k_x)\begin{pmatrix}\alpha_{xx} & 0 \\ 0 & \alpha_{yy}\end{pmatrix}\begin{pmatrix}k_x \\ k_y\end{pmatrix} = \frac{-k_x k_y \alpha_{xx} + k_y k_x \alpha_{yy}}{k_t^2} \\
\langle s|\hat{\alpha}|s\rangle &= \frac{1}{k_t^2}(-k_y \quad k_x)\begin{pmatrix}\alpha_{xx} & 0 \\ 0 & \alpha_{yy}\end{pmatrix}\begin{pmatrix}-k_y \\ k_x\end{pmatrix} = \frac{k_y^2 \alpha_{xx} + k_x^2 \alpha_{yy}}{k_t^2}
\end{aligned}\right\} \quad (19)$$

Equations in (19) indicate that the matrix $\widehat{M}$ is symmetric i.e. $M_{sp} = M_{ps}$. Utilising the explicit form of the matrix elements, the dispersion relation can be transformed to:

$$\left[k_x^2 \alpha_{xx} + k_y^2 \alpha_{yy} + \frac{k_0 k_t^2}{2}\left(\frac{\varepsilon_1}{k_{z1}} + \frac{\varepsilon_2}{k_{z2}}\right)\right]\left[k_y^2 \alpha_{xx} + k_x^2 \alpha_{yy} + \frac{k_t^2}{2k_0}(k_{z1} + k_{z2})\right] -$$

$$k_x^2 k_y^2 (\alpha_{xx} - \alpha_{yy})^2 = 0 \quad (20)$$

Utilising normalized wavevector components (dividing it by $k_0^4$); the Equation (20) can be rewritten as;

$$\left[q_x^2\alpha_{xx} + q_y^2\alpha_{yy} + \frac{q^2}{2}\left(\frac{\varepsilon_1}{q_{z1}} + \frac{\varepsilon_2}{q_{z2}}\right)\right]\left[q_y^2\alpha_{xx} + q_x^2\alpha_{yy} + \frac{q^2}{2}(q_{z1} + q_{z2})\right] -$$

$$q_x^2 q_y^2(\alpha_{xx} - \alpha_{yy})^2 = 0 \quad (21)$$

here $q^2 = q_x^2 + q_y^2$. When $\varepsilon_1 = \varepsilon_2 = 1$ in case of the symmetric vacuum environment, i.e., Equation (21) can be simplified to:

$$k_z k_0(1 + \alpha_{xx}\alpha_{yy}) - (k_x^2\alpha_{xx} + k_y^2\alpha_{yy}) + k_0^2(\alpha_{xx} + \alpha_{yy}) = 0 \quad (22)$$

here $k_z = \sqrt{k_0^2 - k_x^2 - k_y^2}$. The dispersion relation can be generally used to theoretically explain the experimental results and to determine the unknown permittivities such as anisotropic polaritons in α-MoO$_3$ in this case where equation (20) with *α$_{xx}$* and *α$_{yy}$* as fitting parameters, can be used to corroborate the elliptical and hyperbolic features, theoretically.

### X.5 Hyperbolic metasurfaces

Metamaterials are artificially designed materials in order to attain the required properties and functionalities. Metamaterials based on isotropic materials can also be used to efficiently achieve highly large wave confinement and local density of states. In-plane hyperbolic polaritons can be induced in these artificially structured materials Utilising artificially created metallic gratings, the dispersion of surface plasmons polaritons can easily be modified.[77] The periodic shaped metallic gratings could generate surface polaritons with hyperbolic wavefront and orientational selectivity. This can be instantly verified in experimental work based on the hyperbolic metasurface of artificially created silver/air gratings at visible frequency range. [78]

Another example for a hyperbolic metasurface is the utlisation of metallic arrays at microwave frequencies.[79] Using nanostructured graphene, J. S. Gomez-Diaz *et al.*

reported the realization of hyperbolic metasurface at terahertz range.[80] Another recent example of hyperbolic metasurface is the fabrictaion of nanostructured hBN, which can support highly confined anisotropic phonon polaritons (see Figure X.2).[25] These highly confined hyperbolic phonons polaritons demonstrate low losses as compared to plasmon polaritons. We can also develop different hyperbolic metasurfaces at some frequency ranges by suitable engineering of the "meta-atoms".[81]

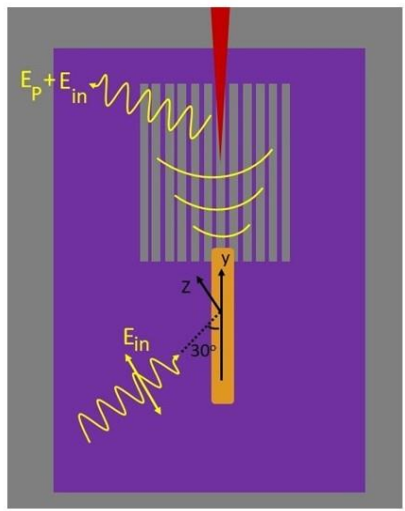

Figure X. 2 Wavefront imaging schematic of antenna-launched hyperbolic metasurface –phonon polaritons

In the context of all these examples, however; their associated losses and undesirable confinement at long wavelengths ($k_p \approx k_0$) can result in a poor performance of optical devices. Moreover, the losses appeared due to fabrication imperfections are also unavoidable. Generally, the maximum momentum of polariton propagation can be associated to the inverse size of "meta-atom", which means the maximum momentum values can halt the device operation. Therefore, their complete potential cannot be utilised due to above mentioned constraints. A recent ground-breaking discovery has demonstrated anisotropic polariton propagation along the surface of a natural vdW α-MoO$_3$ flake (without nanopatterning etc) can enable phonon polariton with elliptic and hyperbolic in-plane dispersion, which are of great interest to infrared optics and nanophotonics.[26] The natural α-MoO$_3$ and BP have demonstrated extreme in-plane optical anisotropies and therefore, hold great potential for directional control of the polaritons.

**X.6 Naturally in-plane anisotropic polaritons**

Since the discovery of graphene in 2004 by Prof Andre Geim and Prof Kostya Novoselov at The University of Manchester (United Kingdom), the broad research area of 2D materials has expanded rapidly in recent years with variety of applications in material science & engineering, chemistry, physics, biological & health sciences etc.[82] With advancement in near field imaging techniques, we are also able to confine light at nanoscale by utilizing 2D materials. Among the broad range of available 2D materials, the in-plane anisotropic materials are the most desirable in nanophotonics applications. The existence of naturally in-plane anisotropy could be related to their orthorhombic structures, which is the most striking characteristic of such natural 2D materials. In this section, we will discuss the progress in the context of in-plane anisotropic polaritons in natural anisotropic materials with a special focus on α-MoO$_3$ and BP. The highly directional propagation of in-plane anisotropic polaritons can be realised because of the presence of large in-plane optical anisotropy in few unique 2D materials. As is well known, optical anisotropy is strongly dependent on the crystal structures and its large values are existing in structurally anisotropic crystals such as biaxial crystals with orthorhombic structure can support large birefringence or dichroism values. Both α-MoO$_3$ and BP crystals have an orthorhombic layered structure and therefore; are exciting materials to investigate for anisotropic optical properties.

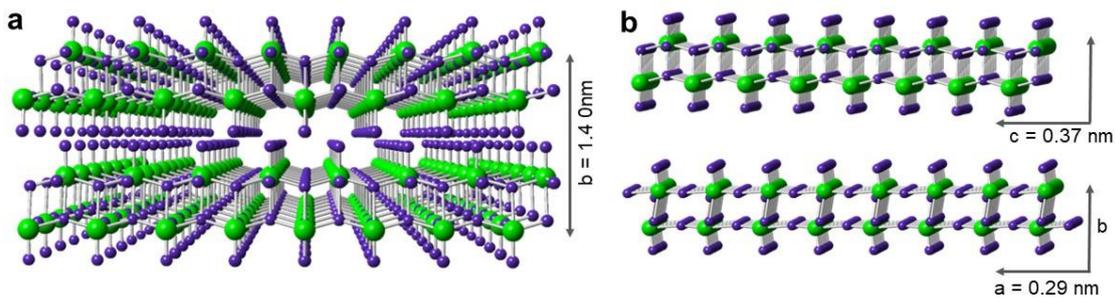

Figure X. 3 Schematic of crystal structure of α-MoO$_3$ (a) Top view of orthorhombic lattice structure in layered (green spheres: molybdenum; blue spheres: oxygen). The lattice constant along b direction is 1.40 nm. (b) in-plane side views of α-MoO$_3$. The lattice constants values are a=0.39 nm, and c=0.37 nm.

Figure X.3a-b shows the schematic diagram for a crystal structure (top and side views) of α-MoO$_3$. The lattice constants (*a*, *b*, *c*) along (*x*, *y*, *z*) directions are all different in real space. The formation of layers is due to distorted edge shared MoO$_6$ octahedra linked by the presence of weakly bonds (i.e. vdW forces). The MoO$_6$ forms corner-sharing chains and edge-sharing chains along the [001] and [100] directions respectively in each double layer structure. As three O atoms are differently linked to a Mo atom, the O$_3$-Mo and O$_2$-Mo are along the [100] and [001] directions respectively. Large in plane anisotropy is existing in α-MoO$_3$ structures because the difference of the spacing between (100) facets and (001) facets is around 7.2%. Similarly, BP atomic layers are stacked together with vdW forces where a single layer is bonded by sp$^3$ phosphorus atoms (see Figure X.4). The obtained layer-to-layer spacing in BP is around 0.52 nm. The BP structure indicates a large in-plane anisotropy values between the crystal orientations along "zigzag" and "armchair" directions as can be seen side-view from different in-plane perspectives (see Figure X.4b).

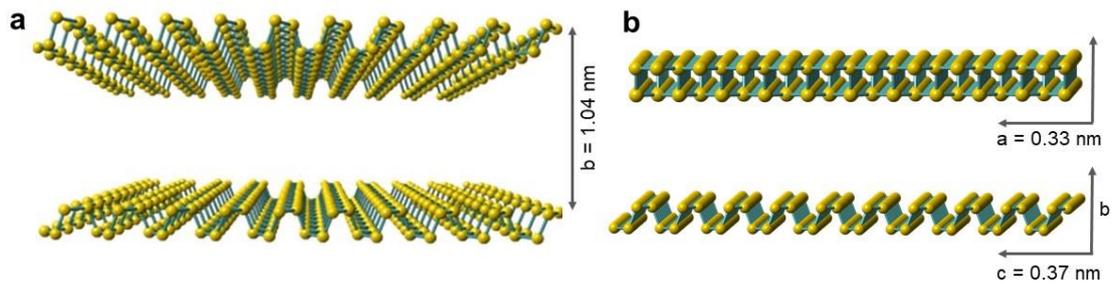

Figure X.4 Schematic of crystal structure of BP (a) Top view of orthorhombic lattice structure. The lattice constant along b direction is 1.04 nm. (b) in-plane side views of BP. The lattice constants values are a=0.33 nm, and c=0.37 nm.

The crystal structures of both $\alpha$-MoO$_3$ and BP crystals have noticeable differences along the two orthogonal directions in the plane. As a result, the obtained dielectric function is highly in-plane anisotropic due to the contributions from the variations in the real part of permittivity values considering as a function of in-plane directions. At resonance frequencies, the strong in-plane anisotropy of permittivity can contribute to directional propagation of polaritons. Therefore; this can indicate that both natural $\alpha$-MoO$_3$ and BP crystals are ideal candidates to investigate in-plane anisotropic polaritons which can lead us to realize the natural hyperbolic surface. There are also many theoretical works which can predict the in-plane anisotropic polaritons in natural materials.[83, 84] These in-plane anisotropic polaritons can be evidenced by the different collective oscillations of electrons or optical phonons propagating along the different directions.

It is worth mentioning here that the mechanisms of in-plane optical anisotropies in both BP and $\alpha$-MoO$_3$ crystals are somehow different in nature. Generally, anisotropy conductivity in case of BP will be considered to describe anisotropic polaritons.[84] The relationship between interband and intraband motions can contribute to different oscillations of electrons or plasmonic oscillations, commonly known as in-plane anisotropic plasmon polaritons. The nanostructure based on metallic materials, oftenly termed as an antenna is normally used to launch the polaritons in desired materials. When it will be illuminated by a desired wavelength, the polaritons propagate away from the antenna in the shape of circular waves like when we throw a stone in a pond, a nice circular wave emerges on the water surface for example hBN[85].

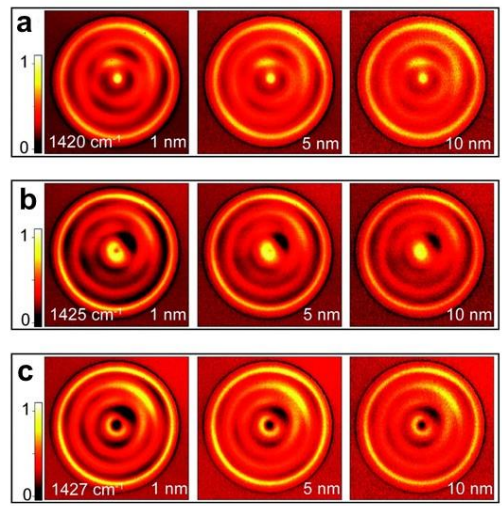

Figure X.5 Fringe patterns of the h-11BN disk through near-field analysis. (b–d) Normalized tomographic PF-SNOM images at tip–sample distances d = 1, 5 and 10 nm (in column) and at ω = 1420, 1425 and 1427 cm−1 (in row)[85].

In the case of BP, the optical responses (launched by an antenna) along both in-plane axes enables the hyperbolic propagation.[24] while controlling the anisotropic conductivity, we can obtain different polaritonic isofrequency contours when a z-oriented dipole emitter is utilised to induce the polaritons along homogeneous surface. The hyperbolic plasmons and topological transitions (from closed elliptical to open hyperbolic) can also be observed over the planar surface. The plasmon propagation along a certain direction can be completely forbidden in case of highly anisotropic regime (Im $\sigma_x * \sigma_y < 0$).[86] Also, their dispersion relation indicates that the resonance modes can only be observed along *x* axis in specific frequency regime above 0.47 ev in this case. The opposite signs of conductivities or $\sigma_{xx}\sigma_{yy} < 0$, is a signature for in-plane hyperbolic plasmon propagation in BP. The imaginary part of BP conductivity could be changed by several approaches such as optical and chemical doping etc in order to tune the in-plane anisotropy.[87, 88] Even though the BP shows a great potential in order to demonstrate the in-plane anisotropic polaritons however, its experimental realization remained elusive so far, because of the unavailability of the suitable mid/far-IR lasers, which will be used as an excitation light source to launch polaritons in real space.

On the other side, the mechanism related to optical in-plane anisotropy can be revealed by highly discrete optical phonons. For polar semiconductor materials such as SiC and *α*-MoO$_3$, the spectral ranges in which anisotropic phonon polaritons exist, are assigned to "Reststrahlen bands" (RBs), which are associated to a long-wavelength incident optical field in the resonance frequency range. The strength of resonances is always dependent on the polarization of the excitation beam. Take *α*-MoO$_3$ as an example; which is a polar semiconductor with vdW layers, is a potential candidate for nanoscale light-confinement because of the presence of strong birefringence.

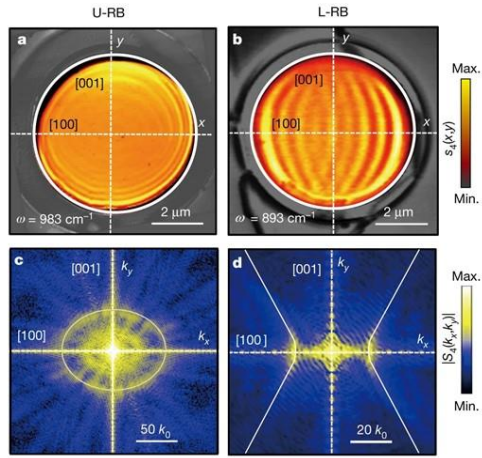

Figure X.6 Near-field real space imaging of polaritons in natural α-MoO$_3$.[26] (a) s-SNOM image of in-plane elliptical phonons polaritons in a disk in upper Reststrahlen Band. b, s-SNOM image of in-plane hyperbolic phonon polaritons in an α-MoO$_3$ disk in lower Reststrahlen Band. (c) Fourier transformation of image in (a). (d) Fourier transformation of image in (b). Both (c) and (d) shows the iso-frequency contours of ellipse and hyperbola for each Reststrahlen Band. Reproduced with permission from ref. 26 Copyright, Nature Publishing Group.

In 2018, Ma and colleagues experimentally discovered the ultralow-loss hyperbolic and elliptical polaritons in natural α-MoO$_3$ flake (see Figure X.6a-d).[26] They demonstrated in-plane hyperbolic phonon polaritons due to in-plane anisotropic dispersion, in natural vdW material. Using s-SNOM, the polariton interferometry real space images reveal highly directional propagation. It is exciting to see the interference pattern in the upper- Reststrahlen Band (U-RB) shows an elliptical shape with the highest phonon polaritons wavelength along the [001] surface direction. On the other side, it transformed into an almond shape in the lower- Reststrahlen Band (L-RB) with the largest wavelength along the [100] surface direction. The Fourier transformation of Figures X.9a and b shows an ellipsoid in the U-RB (Figure X.6c) and hyperbola in the L-RB (Figure X.6d), demonstrating phonons polaritons with elliptic and hyperbolic dispersions, respectively. Among the other discovered polaritons, highest amplitude lifetimes (up to 20 pico-seconds), have also been reported in natural α-MoO$_3$ crystals, which would be useful in technologies based on directional modulation of ultra-fast light-matter interactions (usually at femto-seconds or pico-seconds range).

Unlike the polaritonic propagation in graphene and hBN, the wavefront of the in-plane hyperbolic polaritons in α-MoO$_3$ shows the specific shape in sharp contrast to concentric rings. These anomalous (concave) wavefronts can also be launched by a silver antenna on top of the α-MoO$_3$ flake.[89] The observed wavefronts are emerged due to the presence of strong optical anisotropy in α-MoO$_3$. In natural barium titanium sulfide (BaTiS$_3$), the huge optical in-plane anisotropy was also reported however further work is required to experimentally demonstrate the in-plane anisotropic polaritons.[90] The in-plane anisotropic polaritons in natural vdW crystals reveal the unique characteristics of hyperbolicity in the field of metamaterials, creating an opportunity to develop future technologies based on exotic properties such as such as negative refraction, hyperlens, and enhancement of quantum radiation. Notably, the in plane anisotropic polaritons are observed in a planar surface rather than complicated structures, therefore rendering easier accessibilities.

Besides, the anisotropic plasmon polaritons have still not been experimentally demonstrated in BP. Its anisotropic dispersion can play a role to modulate the in-plane polaritons in hBN by fabricating vertical heterostructures consists of BP and hBN. On another note, low-loss and volume-bound nature of hyperbolic phonon polariton modes in hBN are extensively investigated. In hBN sheet, the phonon polaritons can generally propagate as a cylindrical wave or in-plane isotropic characteristics. The propagation of phonon polaritons in hBN can be modified by stacking hBN disk on top of a BP nanosheet. The anisotropic polaritonic modes can be seen along zigzag and armchair axes of BP.[91] To be specific, the real space nano-images reveal characteristic elliptical interferometry pattern in hBN/BP heterostructure, whereas; a concentric pattern (like Figure X.5) can be seen in pure hBN. These results suggest that fabricating a unique metasurface by stacking vdW materials with different optical birefringence can help us to achieve the tailored anisotropic optical properties. This can also be achieved by longitudinal optical phonon-plasmon coupling effect.[92]

**X.7 Conclusion**

In conclusion, the recent progress of the anisotropic polaritons in vdW crystals have been reviewed in detail. We have obviously focused most of our efforts on paying special attention to the natural vdW materials with in-plane optical anisotropies, which enables polaritons with in-plane anisotropic dispersion. The in-plane optical anisotropies which are linked to the interplay between anisotropic electron motions can generally support the anisotropic surface plasmons and this could also be true for other vdW crystals with similar structures like BP. Few examples are Tellurides and Sulphides.[93-96] In terms of phonon polaritons, the crystals such as $V_2O_5$[97] and $WO_3$[98] which are similar to α-$MoO_3$, are ideal materials to realise anisotropic propagation of polaritons. External parameters including pressure[99-103], doping[39, 104-116] etc which are commonly used to tune the materials physical properties, can be employed to polaritonic field to gain further understanding of underlying physical phenomena, which would be helpful to develop state of the art nanophotonic technologies.

In comparison to conventional metasurfaces, the naturally anisotropic materials are not bound to the size of "meta-atom" and hence illustrates the maximum momentum of polaritons that can be supported. In a more general sense, the natural hyperbolicity could induce an extremely small wavelengths and this can contribute to miniaturizing nanophotonic devices for applications such as sensing, signal processing and quantum computing. Unlike metamaterials, the utilisation of natural materials in polaritonic area can avoid time-consuming fabrication procedures and greatly reduce the related high inherent losses appeared while fabrication processes.

The in-plane anisotropic polaritons in natural vdW materials can help to achieve the planar and directional light modulation with on-chip integration. Even though, the in-plane anisotropic polaritons in natural materials is evidently an exciting but emerging research area due to their peculiar role in future technologies based on mechanisms such as waveguiding, planar negative refraction, hyper-lensing, wavefronts etc. Finally, the in-plane anisotropic polaritons in natural vdW materials has opened unique possibilities for energy control and assist directional heat transfer at nanoscale with extreme efficiencies.

## X.8 References